\documentclass{aastex6}

\newcommand{\vvec}{\mbox{\boldmath$v$}}
\newcommand{\vF}{\mbox{\boldmath$F$}}

\begin{document}

\title{Formulas for Radial Transport in Protoplanetary Disks} 

\author{Steven J. Desch\altaffilmark{1}}
\affil{School of Earth and Space Exploration \\
Arizona State University \\
PO Box 871404, Tempe AZ 85287-1404, USA}
\altaffiltext{1}{\tt steve.desch@asu.edu}

\author{Paul R. Estrada\altaffilmark{2}}
\affil{Space Sciences Division \\
NASA Ames Research Center \\
MS 245-3, Moffett Field, CA, 94035, USA} 

\author{Anusha Kalyaan\altaffilmark{1}}
\affil{School of Earth and Space Exploration \\
Arizona State University \\
PO Box 871404, Tempe AZ 85287-1404, USA}

\and

\author{Jeffrey N. Cuzzi\altaffilmark{2}}
\affil{Space Sciences Division \\
NASA Ames Research Center \\
MS 245-3, Moffett Field, CA, 94035, USA} 

\begin{abstract}
Quantification of the radial transport of gaseous species and solid particles 
is important to many applications in protoplanetary disk evolution.
An especially important example is determining the location of the water snow 
lines in a disk, which requires computing the rates of outward radial diffusion 
of water vapor and the inward radial drift of icy particles;
however, the application is generalized to evaporation fronts of all volatiles.
We review the relevant formulas using a uniform formalism.
This uniform treatment is necessary because the literature currently contains at 
least six  mutually exclusive treatments of radial diffusion of gas, only one of 
which is correct. 
We derive the radial diffusion equations from first principles, using Fick's law.
For completeness, we also present the equations for radial transport of particles.
These equations may be applied to studies of diffusion of gases and particles in 
protoplanetary and other accretion disks.
\end{abstract}


% \end{frontmatter} 

\section{Introduction}

One of the outstanding issues in studies of protoplanetary disks is the issue of radial mixing, both
of gaseous species and particles with a range of sizes from microns to meters. 
Just a few examples of problems illustrate the importance of quantifying such radial diffusion. 
% Planets clearly form at different locations in the solar system and record different chemical compositions,
% presumably due to variations in chemistry; for example, Earth is approximately 8 wt\% FeO while Mars is 
% approximately 18wt\% FeO, reflecting a difference in the chemistry or accretion of volatiles (Rubie et al.\ 2004) 
The {\it Stardust} mission discovered high-temperature condensates resembling fragments of chondrules and
calcium-rich, aluminum-rich
inclusions (CAIs), objects that formed at high temperatures indicative of an origin in the inner solar system,
in the sample return from comet Wild 2, which must have formed in the outer solar system
(Zolensky et al.\ 2006). 
Oxygen isotopic anomalies in meteorites are quite possibly carried by isotopically distinctive water formed
in the outer solar system, radially advected inward as icy particles to the asteroid belt region 
% (Yurimoto \& Kuramoto 2004)
(Lyons et al.\ 2009). 
There is evidence that enstatite chondrites record formation in a part of the solar nebula with elevated 
sulfur composition, possibly related to a sulfur snow line in which sulfur vapor diffuses outward beyond a 
condensation front (Pasek et al.\ 2005; Petaev et al. 2011). 
Finally, ${\rm H}_{2}{\rm O}$ snow lines are very important in our solar system for determining the water content
of asteroids and planets. 
Just beyond the snow lines, outwardly diffusing water vapor can be cold-trapped as ice, enhancing the density of solid 
ice there, possibly triggering Jupiter's formation (Stevenson \& Lunine 1988; Cuzzi \& Zahnle 2004).
In each of these problems it is important to quantify how water or other vapor diffuses relative to the gas, as
well as how particles of various sizes drift and diffuse relative to the gas.

The flow of matter through the disk is described by a formula for the time evolution of the surface density 
of gas, $\Sigma(r,t)$, which is a function of heliocentric distance $r$ and time $t$.
This evolution can be written as two first-order differential equations for the viscous evolution of an 
accretion disk (Lynden-Bell \& Pringle 1974).
The first one describes the change in surface density $\Sigma$ as mass flows into and out of each annulus:
\begin{equation}
\frac{\partial \Sigma}{\partial t} = \frac{1}{2\pi r} \,
 \frac{\partial \dot{M}}{\partial r} = -\frac{1}{r} \, \frac{\partial}{\partial r} \left( r \, \Sigma \, V_{r} \right),
\end{equation}
where $V_{r} = -\dot{M} / (2\pi r)$ is the average gas velocity (positive if outward). 
The second one describes the flow of mass due to the exchange of angular momentum between two adjacent annuli, 
coupled by a viscous torque mediated by turbulent viscosity $\nu$: 
\begin{equation}
\dot{M} = 6\pi \, r^{1/2} \, \frac{\partial}{\partial r} \left(
 r^{1/2} \, \Sigma \nu \right) = 3\pi \Sigma \nu \left( 1 + 2 Q \right),
\end{equation}
where $Q \equiv \partial \ln (\Sigma \nu) / \partial \ln r$ and $\dot{M} > 0$ if the flow is inward.
An important quantity is the radial velocity of the gas,
\begin{equation} 
V_{r} = -\frac{ \dot{M} }{ 2\pi r \Sigma } = -\frac{3\nu}{2r} \, \left( 1 + 2 Q \right).
\end{equation}
(Inward gas flow has $V_{r} < 0$.) 
Boundary conditions placed on $\dot{M}$ at inner and outer
edges of the disk suffice to close the equations.
One can also write these equations as a single second-order differential equation for $\Sigma$:
\begin{equation}
\frac{\partial \Sigma}{\partial t} = \frac{3}{r} \, \frac{\partial}{\partial r} \, 
 \left[ r^{1/2} \, \frac{\partial}{\partial r} \left( r^{1/2} \, \Sigma \nu \right) \right]. 
\end{equation} 
Self-similar solutions have been presented by Lynden-Bell \& Pringle (1974) 
and Hartmann et al.\ (1998), and this treatment is standard in the literature. 

What is sought is an additional formula that describes the evolution of the
surface density of a tracer volatile, $\Sigma_{\rm c}$, or the evolution of the concentration of 
the volatile, $c \equiv \Sigma_{\rm c} / \Sigma$. 
Sometimes the volatiles are in the form of gas (vapor), or sometimes in the form of solid particles
(e.g., icy grains) of potentially any size. 
In Section 2 we show that the literature currently contains multiple mutually exclusive formulas
for the radial transport of gaseous tracer species, which are defined to be dynamically well coupled 
to the gas. 
In Section 3 we derive the correct formula from first principles using Fick's law.
In Section 4, for completeness, we provide a similar formula for the radial transport of solid particles. 
Finally, in section 5 we discuss the different outcomes predicted by various treatments, to illustrate
the importance of using the correct formula. 
Our goal is to clear up the discrepancies in the literature and provide a resource for other researchers 
studying radial transport in accretion disks.

\section{Existing Treatments of Gas Diffusion}

Reviewing the literature, we found at least nine separate, original derivations that include six mutually exclusive 
differential equations describing how the concentration $c$ of a gaseous tracer species changes 
in time due to diffusion relative to the main gas in a protoplanetary disk.
What defines a gaseous tracer species is that it is dynamically well coupled to the gas.
This definition includes gaseous vapor, but could also apply to very small particles, as well. 
Here we review the various formulas for radial transport of gaseous tracer species, 
using a standardized formalism in which the radial velocity 
of the main gas is $V_{r}$, due to a turbulent viscosity $\nu$, and in which the 
tracer species diffuses relative to the gas with diffusion coefficient ${\cal D}_{\rm g}$.
The mass diffusion coefficient of gas, ${\cal D}_{\rm g}$, and the kinematic viscosity $\nu$ are related but 
need not be identical; their ratio is the Schmidt number ${\rm Sc} = \nu / {\cal D}_{\rm g}$.
We consider ${\cal D}_{\rm g}$ and $\nu$ to both vary with heliocentric distance. 

\subsection{Clarke \& Pringle (1988), Gail (2001), Bockel\'{e}e-Morvan et al.\ (2002)}

Clarke \& Pringle (1988), Gail (2001), and Bockel\'{e}e-Morvan et al.\ (2002) each studied the problem of 
radial mixing in the same manner, apparently independently.
Each started with the following equation, given by Morfill \& V\"{o}lk (1984): 
\begin{equation}
\frac{\partial (c \rho)}{\partial t} + \nabla \cdot \left( 
c \, \rho \, \vvec \right) = \nabla \cdot \left( \rho {\cal D}_{\rm g} \,
\nabla c \right)
\end{equation} 
(Gail 2001 cited Hirschfelder et al.\ 1964).
Assuming axisymmetry, integrating this over all $z$, assuming $\rho(z)$ vanishes far from the midplane, and 
assuming $c$ and ${\cal D}_{\rm g}$ are vertically uniform, one derives:
%\begin{equation}
%\frac{\partial \Sigma_{\rm c}}{\partial t} 
% + \frac{1}{r} \frac{\partial}{\partial r} \left( r \, \Sigma_{\rm c} \, V_{r} \right) = 
% + \frac{1}{r} \frac{\partial}{\partial r} \left( 
% r \int_{z=-\infty}^{+\infty} \rho \, {\cal D}_{\rm g} \, \frac{\partial c}{\partial r} \, dz \right) .
%\end{equation} 
%If one now assumes $c$ and ${\cal D}_{\rm g}$ are vertically uniform, one can simplify this as
\begin{equation}
\frac{\partial \Sigma_{\rm c}}{\partial t} 
 + \frac{1}{r} \frac{\partial}{\partial r} \left( r \, \Sigma_{\rm c} \, V_{r} \right) = 
   \frac{1}{r} \frac{\partial}{\partial r} \left( r \, \Sigma \, {\cal D}_{\rm g} \, 
\frac{\partial c}{\partial r} \right) .
\end{equation} 
Likewise, one can vertically integrate the continuity equation, 
\begin{equation}
\frac{\partial \rho}{\partial t} + \nabla \cdot \left( \rho \vvec \right) = 0,
\end{equation}
to find
\begin{equation}
\frac{\partial \Sigma}{\partial t} 
 + \frac{1}{r} \frac{\partial}{\partial r} \left( r \, \Sigma \, V_{r} \right) = 0.
\end{equation} 
Multiplying this equation by $c$ and subtracting from the previous one yields
%\begin{equation}
%\Sigma \, \frac{\partial c}{\partial t} + \Sigma \, V_{r} \, \frac{\partial c}{\partial r} = 
% \frac{1}{r} \frac{\partial}{\partial r} \left( r \, \Sigma \, {\cal D}_{\rm g} \, 
%\frac{\partial c}{\partial r} \right) 
%\end{equation}
%or, finally,
\begin{equation}
\frac{\partial c}{\partial t} +
V_{\rm r} \, \frac{\partial c}{\partial r} =
\frac{1}{r \, \Sigma}
\frac{\partial}{\partial r} \left( r \, \Sigma \, {\cal D}_{\rm g} \, 
\frac{\partial c}{\partial r} \right),
\end{equation}
which is equivalent to equation 2.1.4 of Clarke \& Pringle (1988), equation 12 of Gail (2001), 
and equation 4 of Bockel\'{e}e-Morvan et al.\ (2002).
We note that it is equivalent also to equation 21 of Cuzzi et al.\ (2003), who based their derivation 
on that of Bockel\'{e}e-Morvan et al.\ (2002).
%Only if the product $(\Sigma \, {\cal D}_{\rm g})$ were uniform in $r$ ($Q = 0$) would this would match
%the equation of Stevenson \& Lunine (1988).
If one assumes ${\rm Sc} = 1$, one can write
\begin{equation}
\frac{\partial c}{\partial t} = 
\nu \left[ ( \frac{5}{2} + 4 Q) \,
\frac{1}{r} \frac{\partial c}{\partial r}
+ \frac{\partial^2 c}{\partial r^2} \right].
\end{equation} 

\subsection{Stevenson \& Lunine (1988)}

Stevenson \& Lunine (1988) started with the equation
\begin{equation}
\frac{\partial c}{\partial t} + V_{\rm r} \, \frac{\partial c}{\partial r}
 = \frac{ {\cal D}_{\rm g} }{r} \, \frac{\partial}{\partial r} \left( 
 r \frac{\partial c}{\partial r} \right)
\end{equation}
(their equation 6), 
then argued that $V_{\rm r} \approx -\nu / r \approx -{\cal D}_{\rm g}/r$ on 
dimensional grounds, to derive a simplified formula.
Their treatment captures the physics of the problem but neglects the effects of 
any radial gradients in the density and the diffusion coefficient. 
Assuming ${\rm Sc} = 1$, it is straightforward to show this equation can be rewritten:
\begin{equation}
\frac{\partial c}{\partial t} = 
\nu \left[ ( \frac{5}{2} + 3 Q) \,
\frac{1}{r} \frac{\partial c}{\partial r}
+ \frac{\partial^2 c}{\partial r^2} \right]. 
\end{equation} 

\subsection{Drouart et al.\ (1999)}

A different treatment was adopted by Drouart et al.\ (1999), who wrote:
\begin{equation}
\frac{\partial c}{\partial t} 
+\frac{1}{r} \, \frac{\partial}{\partial r} \, \left( r \, V_{r} \, c \right) = 
 \frac{1}{r \, \Sigma} \, \frac{\partial}{\partial r} \, \left( r \, \Sigma \, {\cal D}_{\rm g}
 \, \frac{\partial c}{\partial r} \right)
\end{equation}
(their equation 9).
This can be rewritten as 
\begin{equation}
\frac{\partial \Sigma_{\rm c}}{\partial t} + \frac{1}{r} \, \frac{\partial}{\partial r} \,  \left( 
 r \, V_{r} \, \Sigma_{\rm c} \right) + \Sigma_{\rm c} \, \frac{1}{r} \, \frac{\partial}{\partial r} \,
 \left( r V_{r} \right) = \frac{1}{r} \, \frac{\partial}{\partial r} \, \left( r \, \Sigma \, {\cal D}_{\rm g}
 \, \frac{\partial c}{\partial r} \right).
\end{equation} 
Assuming ${\rm Sc} = 1$, we can also write this as 
\begin{equation}
\frac{\partial c}{\partial t} = - c \, \left( \frac{V_{r}}{r} + \frac{\partial V_{r}}{\partial r} \right) 
+ \nu \, \left[ \left( \frac{5}{2} + 4 Q \right) \, \frac{1}{r} \frac{\partial c}{\partial r} 
 + \frac{\partial^2 c}{\partial r^2} \right].
\end{equation}
This differs from other treatments by including a term
proportional to the concentration and the divergence of the velocity field, erroneously implying that the 
concentration would increase if the density were to increase.

\subsection{Cuzzi \& Zahnle (2004)}

The treatment of Cuzzi \& Zahnle (2004) is similar but not quite identical. 
Setting $f_{\rm L} = 0$ in their equations 1 and 2 yields
\begin{equation}
\frac{\partial c}{\partial t} = 
V_{{\rm r}} \, \frac{\partial c}{\partial r}
+\frac{1}{r \Sigma} \, \frac{\partial}{\partial r} \, \left[ {\cal D}_{\rm g} \, 
 \frac{\partial}{\partial r} \left( c \Sigma \right) \right].
\end{equation}  
This resembles the equations derived by Clarke \& Pringle (1988) and others, 
but includes several extra terms involving the gradients of 
$\Sigma$ and ${\cal D}_{\rm g}$. 

\subsection{Ciesla \& Cuzzi (2006) and Guillot \& Hueso (2006)}

Two later treatments started with different equations but apparently a common
starting assumption.
Ciesla \& Cuzzi (2006) wrote
\begin{equation}
\frac{\partial \Sigma_{\rm c}}{\partial t} = \frac{3}{r} \,
 \frac{\partial}{\partial r} \left[ r^{1/2} \, \frac{\partial}{\partial r}
 \left( \Sigma_{\rm c} \, \nu \, r^{1/2} \right) \right]
\end{equation}
(their equation 11), 
apparently deriving their equations by assuming $\Sigma_{\rm c}$ would evolve by 
the same differential equation as $\Sigma$.
Guillot \& Hueso (2006) wrote 
\begin{equation}
\frac{\partial c}{\partial t} =3\nu \left[ \left( \frac{3}{2} + 
 2 \frac{\partial \ln (\Sigma \nu)}{\partial \ln r} \right) \,
 \frac{1}{r} \frac{\partial c}{\partial r} + \frac{\partial^2 c}{\partial r^2}
 \right] 
\end{equation}
(their equation 5). 
It is straightforward to show that these are equivalent, and both can be rewritten 
(again assuming ${\rm Sc} = 1$) as 
\begin{equation}
\frac{\partial c}{\partial t} = 
3 \nu \left[ ( \frac{3}{2} + 2 Q) \,
\frac{1}{r} \frac{\partial c}{\partial r}
+ \frac{\partial^2 c}{\partial r^2} \right]. 
\end{equation} 
It is notable that in this treatment the diffusion coefficient is $3 \nu$, not the value $\nu$
that should obtain in the limit of small spatial scales. 

\subsection{Ciesla (2009)} 

Ciesla (2009) derived a diffusion equation starting with the two-dimensional formula
\begin{equation}
\frac{\partial c}{\partial t} + \frac{1}{r \, \rho} \, 
\frac{\partial}{\partial r} \left( r V_{r} \rho c \right) + 
\frac{1}{\rho} \frac{\partial}{\partial z} \left( V_{z} \rho c \right)
= \frac{1}{r \, \rho} \frac{\partial}{\partial r} \left(
r \rho \nu \frac{\partial c}{\partial r} \right) + \frac{1}{\rho}
\frac{\partial}{\partial z} \left( \rho \nu \frac{\partial c}{\partial z} 
\right)
\end{equation}
(his equation 9), 
 which is equivalent to 
 \begin{equation}
 \frac{\partial c}{\partial t} + \frac{1}{\rho} \nabla \cdot
 \left( \rho c \vvec \right) = \frac{1}{\rho} \nabla \cdot
 \left( \rho \nu \nabla c \right),
 \end{equation} 
 assuming axisymmetry. 
In the same manner as before this can be converted to a one-dimensional form, by 
integrating over all $z$, and assuming $\nu$ and $c$ are vertically uniform.
One finds
\begin{equation}
\frac{\partial c}{\partial t} + \frac{1}{r \, \Sigma} 
\frac{\partial}{\partial r} \left( r \, \Sigma \, V_{r} \, c \right)
= \frac{1}{r \, \Sigma} \frac{\partial}{\partial r} \left( r \,
\Sigma \, \nu \, \frac{\partial c}{\partial r} \right).
\end{equation} 
This differs from previous treatments in that the product 
$r \Sigma V_{\rm r}$ is inside the radial derivative on the 
left-hand side. 
This formula can be rewritten as 
%\begin{equation}
%\frac{\partial c}{\partial t} - \frac{1}{2\pi r \, \Sigma} \,
%\frac{\partial}{\partial r} \left( c \dot{M} \right) = 
%\nu \left[ \left( 1 + Q \right) \frac{1}{r} \frac{\partial c}{\partial r} 
% + \frac{\partial^2 c}{\partial r^2} \right].
%\end{equation}
%or
%\begin{equation}
%\frac{\partial c}{\partial t} 
%-\frac{3\pi \, \Sigma \nu \, (1 + 2 Q)}{2\pi r \, \Sigma} \, 
% \frac{\partial c}{\partial r} - \frac{c}{2\pi r \, \Sigma} \,
% \frac{\partial \dot{M}}{\partial r} = 
%\nu \left[ \left( 1 + Q \right) \frac{1}{r} \frac{\partial c}{\partial r} 
% + \frac{\partial^2 c}{\partial r^2} \right],
%\end{equation} 
%or
\begin{equation} 
\frac{\partial c}{\partial t} = \frac{c}{\Sigma} \,
\frac{\partial \Sigma}{\partial t} + \nu \left[ 
\left( \frac{5}{2} + 4 Q \right) \, 
\frac{1}{r} \frac{\partial c}{\partial r} 
 +\frac{\partial^2 c}{\partial r^2} \right].
\end{equation}
This is identical to the formula of Gail (2001) except for the term 
$(c / \Sigma) \, \partial \Sigma / \partial t$ on the left-hand side,
which erroneously assumes that the concentration would change if the 
surface density were to change (as was also assumed by Drouart et al.\ 1999). 

\section{Derivation of Gas Diffusion Equation}

To determine which (if any) of the above equations is correct, we 
derive the volatile radial diffusion equation from first principles. 
A tracer species will be advected with the gas, even as it diffuses relative to it.
We assign an effective mass accretion rate to the species $c$, 
which is the sum of the flux due to its advection in the mean flow,
and the diffusion flux due to concentration gradients, following Fick's law:

\begin{equation}
\frac{\partial}{\partial t} \left( \rho c \right) = -\nabla \cdot \vF,
\end{equation}
where the flux of the tracer species is
\begin{equation}
\vF = \rho c \, \vvec  -{\cal D}_{\rm g} \, \rho \, \nabla c,
\end{equation}
the first term being an advective term and the second term capturing diffusion of 
the tracer relative to the gas.
Again integrating over $z$ and assuming ${\cal D}_{\rm g}$, $c$ and $V_{r}$ are vertically 
uniform, we find 
\begin{equation}
\frac{\partial}{\partial t} \, \left( \Sigma c \right) 
+\frac{1}{r} \, \frac{\partial}{\partial r} \, \left( r \, \Sigma \, c \, V_{r} \right)
= \frac{1}{r} \, \frac{\partial}{\partial r} \, \left( r \, {\cal D}_{\rm g} \, \Sigma \, \frac{\partial c}{\partial r} \right).
\end{equation} 
Clarke \& Pringle (1988) started with this equation [their equation 2.1.4], deriving it from the contaminant equation 
given by Morfill \& V\"{o}lk (1984). 
We believe that Morfill \& V\"{o}lk (1984) are the first to write this (correct) three-dimensional equation in the 
context of protoplanetary disks. 
This equation is equivalent to setting the mass flux of the tracer species to
\begin{equation}
\dot{M}_{\rm c} = c \, \dot{M} 
 +2\pi r \, {\cal D}_{\rm g} \, \Sigma \, \frac{\partial c}{\partial r},
\end{equation}
where ${\cal D}_{\rm g}$ is the diffusion coefficient of species $c$
through the main gas.
The sign of the diffusion flux means that species $c$ diffuses
inward (positive $\dot{M}_{\rm c}$) if $\partial c / \partial r > 0$.
We then write
\begin{equation}
\frac{\partial}{\partial t} \left( c \Sigma \right)
= \frac{1}{2\pi r} \, 
\frac{\partial \dot{M}_{\rm c}}{\partial r},
\end{equation} 
or 
\begin{equation}
c \frac{\partial \Sigma}{\partial t} 
+\Sigma \frac{\partial c}{\partial t} = \frac{1}{2\pi r} \, 
\frac{\partial}{\partial r} \left[ c \dot{M} 
 +2\pi r {\cal D}_{\rm g} \Sigma \frac{\partial c}{\partial r} \right].
\end{equation} 
If we impose $c \equiv 1$ we recover 
\begin{equation}
\frac{\partial \Sigma}{\partial t} 
= \frac{1}{2\pi r} \, 
\frac{\partial \dot{M}}{\partial r},
\end{equation} 
as expected.
Subtracting $c$ times this equation yields 
%\begin{equation}
%\Sigma \frac{\partial c}{\partial t} = 
% \frac{1}{2\pi r} \dot{M} \frac{\partial c}{\partial r} 
%+\frac{1}{2\pi r} \frac{\partial}{\partial r} \left[
% 2\pi r \, {\cal D}_{\rm g} \Sigma \, \frac{\partial c}{\partial r} \right], 
%\end{equation} 
%or 
\begin{equation}
\frac{\partial c}{\partial t} = \frac{1}{2\pi r} \frac{\dot{M}}{\Sigma} \,
 \frac{\partial c}{\partial r} + \frac{1}{r \, \Sigma} 
 \frac{\partial}{\partial r}
 \left[ r \, \Sigma {\cal D}_{\rm g} \, \frac{\partial c}{\partial r} \right].
\end{equation} 
We replace $\dot{M}$ with $3\pi \Sigma \nu (1 + 2 Q)$ to find
\begin{equation} 
\frac{\partial c}{\partial t} = 
\frac{3}{2} \frac{\nu}{r} (1 + 2 Q) \, \frac{\partial c}{\partial r}
+\frac{{\cal D}_{\rm g}}{r} \, (1 + Q') \, \frac{\partial c}{\partial r}
+{\cal D}_{\rm g} \, \frac{\partial^2 c}{\partial r^2},
\end{equation} 
where $Q' \equiv \partial \ln (\Sigma {\cal D}_{\rm g}) / \partial \ln r$.
Substituting ${\cal D}_{\rm g} = \nu / {\rm Sc}$, 
\begin{equation} \label{eq:bigkahuna}
\frac{\partial c}{\partial t} = \nu \left[
 \left( \frac{3}{2} (1 + 2 Q) + \frac{ (1 + Q') }{ {\rm Sc} } \right)
 \, \frac{1}{r} \frac{\partial c}{\partial r} + \frac{1}{\rm Sc}
 \, \frac{\partial^2 c}{\partial r^2} \right].
%\label{eq:bigkahuna} 
\end{equation}
Note that the roles of diffusion (dependent on ${\cal D}_{\rm g}$ and advection 
(affected by the gas velocity, and therefore $\nu$) are separated. 
Equation~\ref{eq:bigkahuna} is the proper equation to track the  radial evolution 
of gas-phase volatiles in protoplanetary disks.
We note that it is more general than many existing treatments, and clearly delineates
the effect of the Schmidt number not equal to unity. 

For most normal solar nebula turbulent flows, it is very likely that the Schmidt number
is constant at ${\rm Sc} \approx 0.7$
(e.g., Launder 1976, McComb 1990; Johansen et al.\ 2007; Hughes \& Armitage 2010). 
In the limit ${\rm Sc} = 1$, 
\begin{equation}
\frac{\partial c}{\partial t} = 
\nu \left[ ( \frac{5}{2} + 4 Q) \,
\frac{1}{r} \frac{\partial c}{\partial r}
+ \frac{\partial^2 c}{\partial r^2} \right],
\end{equation} 
which matches the equations derived by Clarke \& Pringle (1988), Gail (2001), and 
Bockel\'{e}e-Morvan et al.\ (2002). 

In summary, we found six different, mutually exclusive equations in the literature to describe 
the diffusion of gas in protoplanetary disks. 
We advocate use of the most general form, Equation~\ref{eq:bigkahuna} above, which
works for arbitrary Schmidt number. 
In the limit ${\rm Sc} = 1$, this equation matches those of Clarke \& Pringle (1988), Gail (2001), 
Bockel\'{e}e-Morvan et al.\ (2002), and Cuzzi et al.\ (2003). 
Even in the limiting case ${\rm Sc} = 1$, other treatments differ.

\section{Radial Diffusion and Drift of Particles}

Calculating the radial transport of solids is just as fundamental a problem as 
calculating the radial transport of gases.
While very small ($\sim$ micron-sized) particles have the same transport properties as
gas, the transport of larger particles is more complicated, because such 
particles not only are advected and diffuse relative to the gas, they also can
drift relative to the gas.
The basis for particle drift is that gas in a protoplanetary disk, being partially 
supported against gravity by a pressure gradient force, orbits the star with a velocity
less than the Keplerian velocity; particles, which try to maintain an orbit at Keplerian 
velocity around the star, feel a headwind that makes them lose angular momentum and 
spiral in toward the star.
For completeness, we discuss various approaches to the calculation of these effects.

The radial transport of particles is governed by the same general formulas as the gas is.
Defining the concentration of solid particles as $c \equiv \Sigma_{\rm c} / \Sigma$, 
one again has 
\begin{equation}
\frac{\partial}{\partial t} \left( c \Sigma \right) = \frac{1}{2\pi r} \, \frac{\partial \dot{M}_{\rm c}}{\partial r},
\end{equation}
as in Equation 27 for the gas, but where the mass accretion rate associated with particles
includes advection, diffusion, and drift.
One approach is to treat particles exactly as a gaseous fluid, with $\dot{M}_{\rm c}$ defined
as Equation 26, but with an additional term for drift at velocity $\delta u$ with respect to the gas, so that
\begin{equation}
\dot{M}_{\rm c} = c \dot{M} - 2\pi r \, c \Sigma \, \left( \Delta u \right) 
 + 2\pi r \, {\cal D}_{\rm p} \, \Sigma \, \frac{\partial c}{\partial r} 
\end{equation}
(here we assume $\Delta u < 0$ if particles drift inward, making $\dot{M}_{\rm c}$ more positive).
Equivalently, 
\begin{equation}
\dot{M}_{\rm c} = -2\pi r \, c \Sigma \, \left[ -V_{r} + \Delta u \right]
 + 2\pi r \, {\cal D}_{\rm p} \, \Sigma \, \frac{\partial c}{\partial r},
\end{equation}
where $V_{r} + \Delta u$ represents the total radial velocity of the particles.

To calculate the total radial velocity of the particles, we favor the approach of Takeuchi \& Lin (2002)
[see also Nakagawa et al.\ (1986), Birnstiel et al.\ (2010), and Estrada et al.\ (2016)], which we reproduce here.
Gas orbits the star with an angular velocity $\Omega = \Omega_{\rm K} \, \left( 1 - \eta \right)^{1/2}$,
where $\eta = -(r \Omega_{\rm K}^{2})^{-1} \rho_{\rm g}^{-1} \, \partial P_{\rm g} / \partial r$ $\sim 10^{-3}$, 
$\rho_{\rm g}$ and $P_{\rm g}$ are the gas density and pressure, and $\Omega_{\rm K}$ the Keplerian 
orbital frequency.
The gas has velocity $V_{{\rm g},\phi} = r \Omega$ in the azimuthal direction and $V_{{\rm g},r}$ in the
radial direction. 
Particles have azimuthal velocity $V_{{\rm p},\phi}$ and radial velocity $V_{{\rm p},r}$ that differ
from the gas velocity, and therefore paticles experience a drag force. 
%In the limit that particles are small (in the Stokes regime), the radial component of the force equation is
%\begin{equation} 
%\frac{d V_{{\rm p},r}}{dt} = \frac{V_{\rm p,\phi}^2}{r} - \Omega_{\rm K}^2 r 
% - \frac{\Omega_{\rm K}}{{\rm St}} \, \left( V_{{\rm p},r} - V_{{\rm g},r} \right).
%\end{equation}
The radial component of the force equation is
\begin{equation} 
\frac{d V_{{\rm p},r}}{dt} = \frac{V_{\rm p,\phi}^2}{r} - \Omega_{\rm K}^2 r 
- \frac{1}{t_{\rm stop}} \, \left( V_{{\rm p},r} - V_{{\rm g},r} \right), 
\end{equation}
where $t_{\rm stop}$ is the aerodynamic stopping time, defined by matching the acceleration from the
drag force to the term above. 
Likewise, particles lose angular momentum due to the aziumthal drag force:
%\begin{equation} 
%\frac{d}{dt} \, \left( r V_{{\rm p},\phi} \right) = 
% - \frac{\Omega_{\rm K}}{{\rm St}} \, \left( V_{{\rm p},\phi} - V_{{\rm g},\phi} \right).
%\end{equation}
\begin{equation} 
\frac{d}{dt} \, \left( r V_{{\rm p},\phi} \right) = 
 - \frac{1}{t_{\rm stop}} \, \left( V_{{\rm p},\phi} - V_{{\rm g},\phi} \right).
\end{equation}
Using $t_{\rm stop}$, we define the Stokes number as the product of orbital frequency and aerodynamic stopping time:
%\begin{equation}
%{\rm St} = \Omega_{\rm K} \, \frac{ \rho_{\rm p} a }{ \rho_{\rm g} C_{\rm s} },
%\end{equation}
\begin{equation}
{\rm St} = \Omega_{\rm K} \, t_{\rm stop}.
\end{equation}
In the special case of particles smaller than the molecular mean free path (Epstein limit), 
the aerodynamic stopping time and Stokes number can be written in terms of particle properties as
\begin{equation}
{\rm St} = \Omega_{\rm K} \, \frac{ \rho_{\rm p} a }{ \rho_{\rm g} C_{\rm s} },
\end{equation}
where $\rho_{\rm p}$ and $a$ are the particle density and radius, and $C_{\rm s}$ the sound speed,
appropriate for particles smaller than the molecular mean free path (Weidenschilling 1977; 
Cuzzi \& Weidenschilling 2006). 
For particles with radius $a = 1 \, {\rm mm}$, at 1 AU, assuming $\rho_{\rm g} = 10^{-9} \, {\rm g} \, {\rm cm}^{-3}$
and $C_{\rm s} \sim 1 \, {\rm km} \, {\rm s}^{-1}$, ${\rm St} \sim 10^{-3}$. 
In the limit of small particles, such that ${\rm St} \ll 1$, Takeuchi \& Lin (2002) find a solution; assuming 
$V_{{\rm g},\phi} \approx V_{{\rm p},\phi} \approx r \, \Omega_{\rm K}$, they find
\begin{equation}
V_{{\rm p},r} = \frac{ V_{{\rm g},r} - \eta \, {\rm St} \, r \, \Omega_{\rm K} }{ 1 + {\rm St}^2 }.
\end{equation}
The drift speed in this case is 
\begin{equation} \label{eq:43}
\Delta u = V_{{\rm p},r} - V_{{\rm g},r} = 
 \frac{ -{\rm St}^2 \, V_{{\rm g},r} - \eta \, {\rm St} \, r \, \Omega_{\rm K} }{ 1 + {\rm St}^2 },
\end{equation} 
which is generally negative (inward), and which vanishes for small particles. 
For millimeter-sized particles, with ${\rm St} \approx 10^{-3}$, the drift speed at 1 AU is 
$\approx \, V_{{\rm g},r} - \eta \, {\rm St} \, r \Omega_{\rm K}$, both terms being of order 
$\sim  10^{-6} \, {\rm AU} \, {\rm yr}^{-1}$.

Weidenschilling (1977) derived a formula for the drift speed starting with the same assumptions,
but valid for particles of all sizes, not just small particles in the Epstein regime.
He showed that meter-sized particles would drift inward very rapidly, at rates
$\sim \, 10^{-2} \, {\rm AU} \, {\rm yr}^{-1}$, orders of magnitude greater than the drift rate
of millimeter-sized particles.
Unfortunately, in the small-particle limit, the calculation of Weidenschilling (1977) does not
reproduce that of Takeuchi \& Lin (2002): when relating the loss of angular momentum to the 
radial velocity of particles (Weidenschilling's Equation 19), it is assumed that particles have
radial velocity $\Delta u$ instead of $\Delta u + V_{{\rm g},r}$.
While this approximation is appropriate for rapidly drifting particles
(or a disk with no radial flow), it is not appropriate for 
small particles for which $\left| \Delta u \right| \, \sim \, \left| V_{{\rm g},r} \right|$.
We therefore prefer the formulation of Takeuchi \& Lin (2002) to account for the particle advection 
and drift.

The last term to modify for radial transport of particles is the diffusion term.
Vapor and very small particles diffuse relative to the gas with diffusion coefficient 
${\cal D}_{\rm g}$ that differs from the turbulent viscosity $\nu$ by a factor equal
to the Schmidt number: ${\cal D}_{\rm g} = \nu \, {\rm Sc}^{-1}$.
Larger particles will diffuse a rate that differs from this value, depending on their
aerodynamic stopping time and the level of turbulence. 
%
%Whatever the value of ${\rm Sc}$ for gases in the nebula, particles and vapors can have different 
%diffusion coefficients, depending on their aerodynamic stopping times and the nature of the turbulence
%(V\"{o}lk et al.\ 1980). 
%For example, even small particles diffuse differently from gas in response to magnetohydrodynamic
%turbulence (Johansen et al.\ 2006), so that 
%${\rm Sc} = \nu / {\cal D}_{\rm p}$ even in the limit ${\rm St} \ll 1$ can differ from unity.
%Larger particles diffuse at still different rates that depend on their sizes, such that 
%
We adopt the relationship 
\begin{equation}
{\cal D}_{\rm p} = \frac{ {\cal D}_{\rm g} }{ 1 + {\rm St}^2 } = \frac{ \nu \, {\rm Sc}^{-1} }{ 1 + {\rm St}^2 }.
\end{equation}
(Youdin \& Lithwick 2007; Carballido et al.\ 2011).
It will be convenient to define a new quantity
\begin{equation}
Q'_{\rm p} \equiv \frac{r}{\Sigma {\cal D}_{\rm p}} \, \frac{\partial}{\partial r} \, \left( \Sigma {\cal D}_{\rm p} \right),
\end{equation}
analogous to the similar quantity $Q'$ involving ${\cal D}$.

Combining the above equations, we derive a differential equation for $c$, the concentration of particles:
\[
\frac{\partial c}{\partial t} = \nu \, \left[ 
 \left( \frac{3}{2} \left( 1 + 2 Q \right) + \frac{ \left( 1 + Q'_{\rm p} \right) }{ {\rm Sc} \, ( 1 + {\rm St}^2 ) } \right) \,
 \frac{1}{r} \frac{\partial c}{\partial r} 
+ \frac{1}{{\rm Sc} ( 1 + {\rm St}^2 ) } \, \frac{\partial^2 c}{\partial r^2} \right] 
-\frac{1}{r \Sigma} \, \frac{\partial}{\partial r} \, \left[ r \, \left( \Delta u \right) \, c \Sigma \right]
\]

Because the drift speed $\Delta u$ can vary in a complicated way with particle size and position in the disk,
we do not attempt to generalize this equation further. 
For a more detailed discussion of the transport of particles with arbitrary stopping times, and the role 
of the Schmidt number, see Estrada et al.\ (2016), whose equation 11 is consistent with our Equations 33 and 46.

Among other treatments in the literature of the radial transport of particles, we find that 
the treatment of Birnstiel et al.\ (2010) is essentially identical to the treatment presented here.
We note that the treatment of Brauer et al.\ (2008) follows essentially the same lines, but assumes the diffusion
coefficient of particles is ${\cal D}_{\rm p} = {\cal D}_{\rm g} \, ( 1 + {\rm St} )^{-1}$, following 
V\"{o}lk et al.\ (1980), Cuzzi et al.\ (1993) and Schr\"{a}pler \& Henning (2004), 
instead of the form we prefer here, following Youdin \& Lithwick (2007) and Carballido et al.\ (2011).
Our treatment differs from that of Stepinski \& Valageas (1996, their equation 18), which resembles the
treatment of Guillot \& Hueso (2002) and Ciesla \& Cuzzi (2006) for gas diffusion, as described above.

\section{Discussion}  

Radial transport of gaseous volatiles and solid particles is a problem that arises often in studies of accretion disks,
especially in studies of snow lines in protoplanetary disks. 
Given the fundamental importance of volatile transport, it is unfortunate that so many discrepant treatments of it exist 
in the literature.
The various formulas appear similar in form, but in fact they can predict quite different results. 
In this section we explore in detail some implications of the use of different volatile transport formulations in modeling 
disk evolution, using a simple $\alpha$ disk model to demonstrate disk behavior under varying turbulent strengths using three 
different volatile treatments. 
In addition, we also use water as our tracer species, and explore the effects that each treatment has 
on the total water abundance across the disk. 

\subsection{Effects of Diffusion in a Uniform $\alpha$-disk}
To illustrate the effects of diffusion, we perform simulations of simple $\alpha$ disks that incorporate three different 
formulations for volatile transport: i) the treatment used in our work [also that of Clarke \& Pringle (1988), Gail (2001), 
and Bockel\'{e}e-Morvan et al.\ (2002)]; ii) the treatment of Guillot \& Hueso (2006) and Ciesla \& Cuzzi (2006) 
[hereafter GH06/CC06]; and iii) Stevenson \& Lunine (1988) [hereafter SL88]. 
The results of the above simulations are shown with the evolution of a dye, initially placed in an annulus between 1 and 2 AU 
at time $t = 0$, at times $t = 0.1 \, {\rm Myr}$ and $t = 1 \, {\rm Myr}$, using the three different formulations. 
Disk simulations are performed using the following values of $\alpha$: 10$^{-4}$, 10$^{-3}$ and 10$^{-2}$. 

The underlying evolution of the disk is determined by assuming $\nu = \alpha c_{\rm s}^2 \Omega^{-1}$, where 
$c^2_{\rm s} = (k T / \bar{m})$ is the sound speed, $T(r) = 100 \, (r / 1 \, {\rm AU})^{-1/2} \, {\rm K}$ the disk 
temperature, $r$ the heliocentric distance, and $\bar{m} = 2.33 \, m_{\rm p}$ the mean molecular weight. 
% A constant value $\alpha = 10^{-3}$ has been assumed. 
%\textbf{The initial surface density is }$\Sigma(r) \approx 6000 \, (r / 1 \, {\rm AU})^{-3/2} \, {\rm g} \, {\rm cm}^{-2}$, as per Kalyaan et al.\ (2015).
The initial surface density is $\Sigma(r) \approx 6000 \, (r / 1 \, {\rm AU})^{-3/2} \, {\rm g} \, {\rm cm}^{-2}$, 
as per Kalyaan et al.\ (2015).
Photoevaporation due to the minimum plausible irradiation by an external ultraviolet field 
($G_0 = 0.1$) is assumed. 
We thereafter numerically calculate the evolution of the disk using the treatment of Kalyaan et al.\ (2015). 
For the alternate formulations tested here, we converted the original $\Sigma_c$ evolution equations of GH06/CC06 and 
SL88 to a mass accretion rate $\dot{M_c}$ for implementation into our code. 
This required us to assume that $(1/r) \partial (\Sigma D)/\partial r = 0 $ (i.e., $Q' = 0$) in order to make the 
differential equation easily solvable. 
Therefore in all cases of the alternative treatments, we use identical $\Sigma D$, fixed at value at 1 AU so that $Q' = 0$.
These equations are as follows.
For GH06/CC06, 
\begin{equation}
\dot{M}_{\rm c} = c \, \dot{M} + 6\pi r \, \left( \Sigma {\cal D} \right) \, \frac{\partial c}{\partial r}; 
% \dot{M_c} = c \dot{M} + 6 \pi r \big( \Sigma D\big)\, \frac{\partial c}{\partial r};
\end{equation}
for SL88, 
\begin{equation}
%\dot{M_c} = c \dot{M} - \pi c \Sigma D + 2\pi r \big(  \Sigma D \big)\, \frac{\partial c}{\partial r};
\dot{M}_{\rm c} = c \, \dot{M} -\pi c \Sigma {\cal D} \, + 2\pi r \, \left( \Sigma {\cal D} \right) \, \frac{\partial c}{\partial r}; 
\end{equation}
and for comparison, the equation we use is
\begin{equation}
%\dot{M_c} = c \dot{M} + 2 \pi r \big( \Sigma D\big)\, \frac{\partial c}{\partial r}.
\dot{M}_{\rm c} = c \, \dot{M} + 2\pi r \, \left( \Sigma {\cal D} \right) \, \frac{\partial c}{\partial r}. 
\end{equation}

Figures 1, 2 and 3 show results of our simulations for intermediate ($\alpha=10^{-3}$), low ($\alpha=10^{-4}$) and high 
($\alpha=10^{-2}$) values of $\alpha$.
Figure 1 ($\alpha = 10^{-3}$) illustrates that the different treatments predict very different volatile concentrations in the 
outer protoplanetary disk for $\alpha=10^{-3}$. At 10 AU, after 0.1 Myr, the concentration should be $3 \times 10^{-8}$, and 
at 1 Myr it should be $8 \times 10^{-4}$. 
The treatment of Stevenson \& Lunine (1988) predicts values of $2 \times 10^{-7}$ and $\approx 1.5 \times 10^{-3}$.
The treatments of Guillot \& Hueso (2006) and Ciesla \& Cuzzi (2006) predict values of 
$2 \times 10^{-4}$ and $\approx 3 \times 10^{-3}$. 
Use of the incorrect equation can overestimate the volatile concentration by orders of magnitude. 
Figure 2 ($\alpha = 10^{-4}$) illustrates that the GH06/CC06 equations show diffusion of vapor is enhanced by several orders 
of magnitude both inward and outward of the annulus, at both 0.1 and 1 Myr. 
At 4 AU, GH06/CC06 predicts $c$ = 1 $\times$ 10$^{-4}$ at 0.1 Myr and 8 $\times$ 10$^{-1}$ at 1 Myr. 
In contrast, SL88 and our work predict similar concentrations of 1 $\times$ 10$^{-8}$ at 0.1 Myr and $\sim$ 4 $\times$ 10$^{-3}$ at 1 Myr.
In Figure 3 ($\alpha = 10^{-2}$), it is seen that high $\alpha$ leads to greater turbulent mixing in disks, leading to largely 
similar profiles, deviating in only factors of a few. 
Our treatment predicts $c$ = 6 $\times$ 10$^{-3}$ at 0.1 Myr, and $\approx$ 3 $\times$ 10$^{-4}$ at 1 Myr, at 1 AU. 
For comparison, GH06/CC06 predict  6 $\times$ 10$^{-3}$ (same as ours) and $\approx$ 2 $\times$ 10$^{-3}$, 
and SL88 predict 1 $\times$  10$^{-2}$ and 1 $\times$ 10$^{-3}$.

As expected, the choice of the treatment for volatile transport is more significant for disks with lower $\alpha$, but
for reasonable values of $\alpha$, pertinent to weakly turbulent midplanes of protoplanetary disks, 
using the correct treatment is clearly critical to accurately predicting volatile transport. 

%-----------------------------------------------------------------------------------------------------------
%
% FIGURE 1
%
\begin{figure}[ht!]
\begin{center}
\includegraphics[width=0.50\paperwidth]{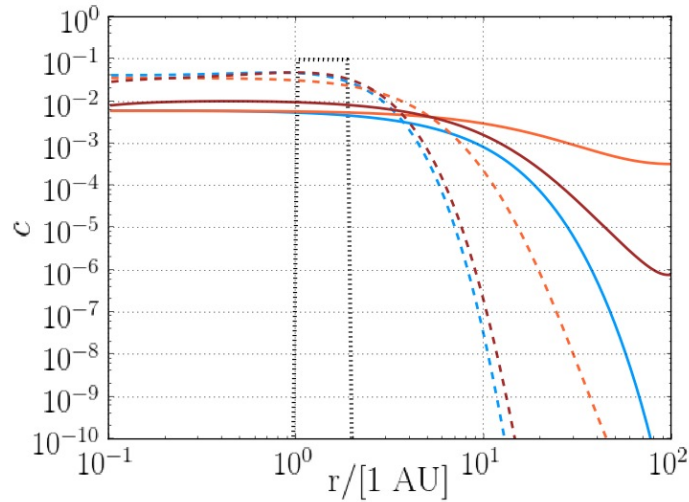}
% \plotone{Fig1.jpg}
\caption{Simulated evolution over time of the concentration, $c(r)$, of a tracer volatile, using three different formulas to describe the radial transport. All three cases begin with the same initial concentration: $c(r) = 0.1$ between 1 and 2 AU (dotted lines) and have turbulent viscosity $\alpha=10^{-3}$. The blue curves show the $c(r)$ at 0.1 Myr (dashed) and 1 Myr (solid) using the formula we suggest (cf. Gail 2001). The orange curves likewise show the evolution using the equation of Guillot \& Hueso (2006) and Ciesla \& Cuzzi (2006), and the brown curves show the evolution using the equation of Stevenson \& Lunine (1988). These two formulations tend to overestimate the amount of diffusion that occurs.} 
\end{center}
\end{figure}
%
%-----------------------------------------------------------------------------------------------------------

%-----------------------------------------------------------------------------------------------------------
%
% FIGURE 2
%
\begin{figure}[ht!]
\begin{center}
\includegraphics[width=0.50\paperwidth]{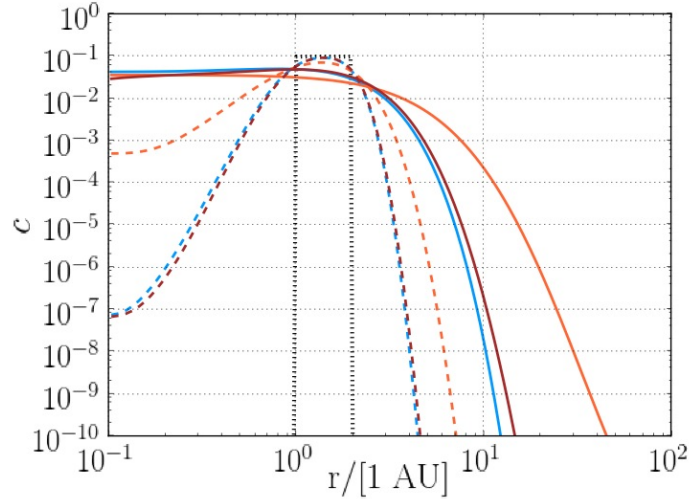}
% \plotone{Fig2.jpg}
\caption{Simulated evolution over time of the concentration, $c(r)$, of a tracer volatile using different 
radial transport formulas.  Same as Figure 1, but for disks with lower turbulent viscosity, $\alpha$ = 10$^{-4}$ . 
GH06/CC06 are seen to overestimate volatile diffusion in disks with low $\alpha$.}
\end{center}
\end{figure}
%
%-----------------------------------------------------------------------------------------------------------

%-----------------------------------------------------------------------------------------------------------
%
% FIGURE 3
%
\begin{figure}[ht!]
\begin{center}
\includegraphics[width=0.50\paperwidth]{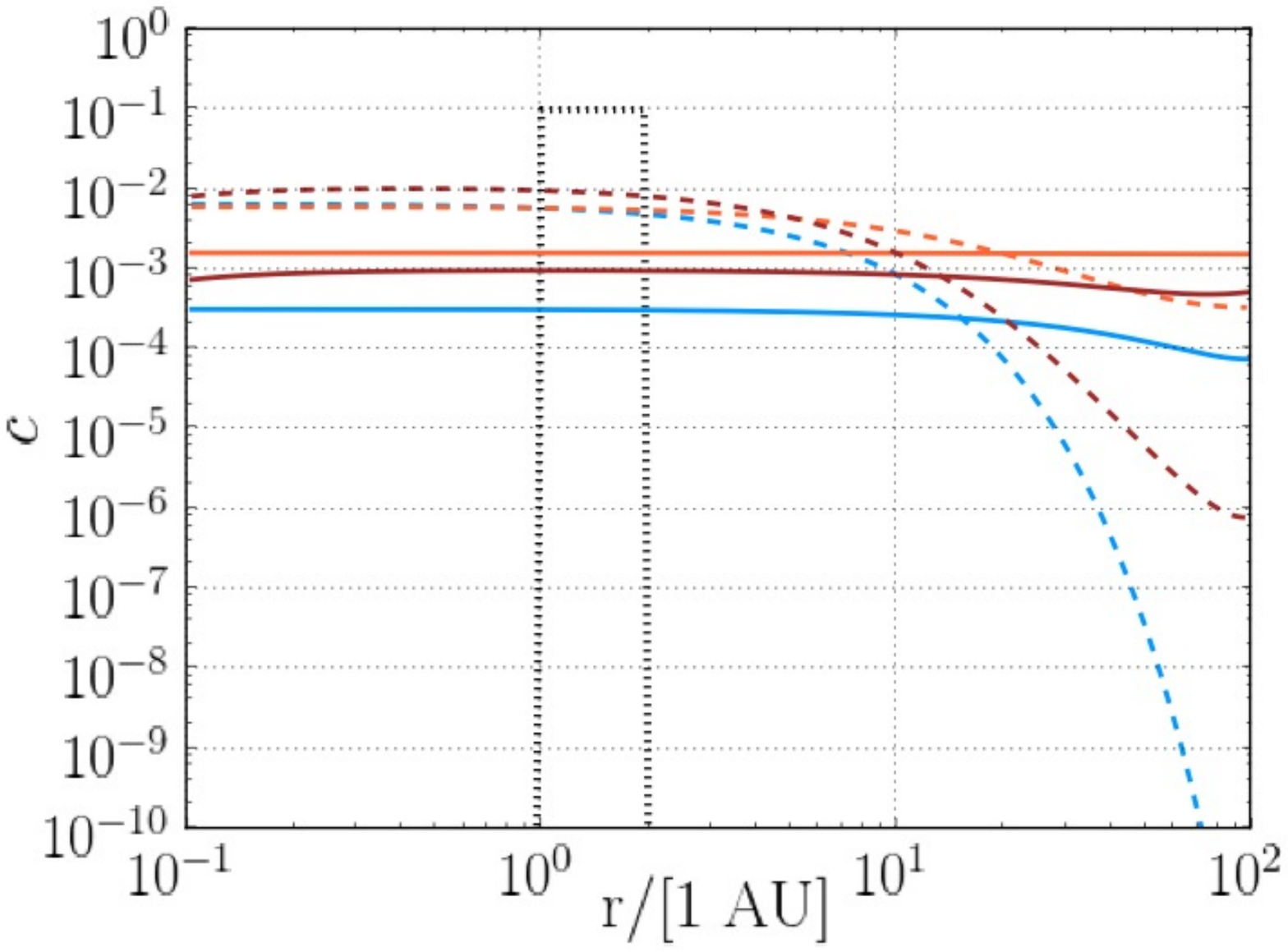}
% \plotone{Fig3.jpg}
\caption{Simulated evolution over time of the concentration, $c(r)$, of a tracer volatile using different 
radial transport formulas.  Same as Figure 1, but for disks with higher turbulent viscosity, $\alpha$ = 10$^{-2}$.
GH06/CC06 and SL88 both predict slightly faster volatile diffusion in disks with high $\alpha$.}
\end{center}
\end{figure}
%
%-----------------------------------------------------------------------------------------------------------

\subsection{Radial Water Abundance Across the Snow Line}

As a second illustration of the effects of the different formulas, for a case with particle transport, we 
build on the above disk model and include both volatile and particle transport,  to determine the water ice-to-rock 
ratio across the disk.
We use our multi-fluid code to track each of the following fluids: 
bulk disk gas ($\Sigma_g$); water vapor ($\Sigma_{\rm vap}$); 
`icy' chondrules composed entirely of water ice ($\Sigma_{\rm icy chondrules}$); 
`rocky' chondrules composed of silicates ($\Sigma_{\rm rocky chondrules}$); 
icy asteroids that grow by accreting icy chondrules ($\Sigma_{\rm icyast}$);
and rocky asteroids that grow by accreting `rocky' chondrules ($\Sigma_{\rm rockyast}$).
Further details will be presented by Kalyaan et al.\ (2017 \textit{in prep.}). 
Uniform tracer concentrations $c(r) \sim$ 10$^{-4}$ for each component are assumed initially throughout the disk. 
Both `icy' and `rocky' chondrules are assumed to be small solid particles of 1 mm diameter and have the same initial 
uniform surface density throughout the disk, $5 \times 10^{-3} \, \Sigma_{\rm gas}$. 

Throughout the disk's evolution, both icy and rocky chondrules radially drift inwards in the disk according to 
Equation~\ref{eq:43}. 
Icy chondrules are additionally influenced by the change of phase of water ice to vapor at pressure and temperature 
conditions close to the snow line region. 
To determine what phase of water exists at a given location, we first use the following equations to determine the saturation water vapor pressure over ice at each radius $r$.
For $T>169\,$K, we use the formulation from Marti \& Mauersberger (1993), 
\begin{equation}
P_{\rm vap}(R) = 0.1 \, \exp \, \left( 28.868 - \frac{6132.9 \, {\rm K}}{T} \right) \, {\rm dyn} \, {\rm cm}^{-2}
\end{equation}
while for $T < 169 \, {\rm K}$ we use the formulation from Mauersberger \& Krankowsky (2003):
\begin{equation}
P_{\rm vap}(R) = 0.1 \, \exp \, \left( 34.262 - \frac{7044.0 \, {\rm K}}{T} \right) \, {\rm dyn} \, {\rm cm}^{-2}.
\end{equation}
We assume that the above formulation for $T\,<\,169\,$K (Mauersberger \& Krankowsky, 2003) is sufficiently accurate when 
extrapolated to lower temperatures ($\sim$150K), as the authors suggest.
The surface density equivalent of $P_{\rm vap}$ is then calculated as
\begin{equation}
\Sigma_{H_2 O, {\rm eq}}(T) = \sqrt{2\pi} \, \left( \frac{ P_{\rm vap} }{ c_s^2} \right).
\end{equation}

The densities of vapor and ice in this condensation-evaporation region are then determined as follows. 
If $\Sigma_{H_2O,{\rm vap}}(T)$ exceeds the total water content (excluding water already accreted into asteroids) at radius $r$ 
(i.e., $\Sigma_{\rm vap}$ $+$ $\Sigma_{\rm icy\,chondrules}$), we assume that all of the water is converted into vapor. 
If on the other hand, $\Sigma_{H_2O,vap}(T)$ is less than the total water content (excluding water in asteroids), then we assume 
that $\Sigma_{\rm vap}$ = $\Sigma_{H_2O,{\rm eq}}$, and the remaining water is in water ice. 
Asteroids are assumed to grow from chondrules at a timescale $t_{\rm growth}$ $\sim$ 1 Myr and behave as a sink for water beyond 
the snow line, as follows:
\begin{equation}
\frac{\partial \Sigma_{\rm icyast}}{\partial t} = \frac{\Sigma_{\rm icy\,chondrules}}{t_{\rm growth}}
\end{equation}
 Radial drift of asteroids and their migration is ignored in this study.

We have incorporated the different diffusion formulations into an $\alpha$-disk model as in \S 5.1, with radial particle transport 
and condensation-evaporation of volatiles as described above, along with accretional heating. 
Following Lesniak \& Desch (2011), this model assumes that accretion is dominant only in the active surface layers of the disk, 
whose surface density is assumed to be $\Sigma_{\rm act}$ $\sim 10 \, {\rm g} \, {\rm cm}^{-2}$, for which the optical depth through 
the active layer is $\tau = \kappa \,\Sigma_{\rm active}$, where $\kappa = 10 \, {\rm cm}^{2} \, {\rm g}^{-1}$. 
Thereafter, the midplane temperature is calculated as follows:
\begin{equation}
\sigma T^{4}_{\rm mid} \approx \sigma T^4_{\rm passive} + \frac{27}{32}\, \Sigma_{\rm act} \nu_{\rm act} \Omega^2\,\tau
\end{equation}
Here, $\nu_{\rm act} = \alpha c_s H$, where $\alpha_{\rm act}$ is assumed to be 0.1. 

Figure 4 traces the distribution of water across the disk, by plotting the total water content 
[i.e., ($\Sigma_{\rm vapor} + \Sigma_{\rm icy chondrules} + \Sigma_{\rm icy asteroids}) / \Sigma_{\rm gas}$] against 
heliocentric distance $r$, at 0.1 Myr (dashed) and 1 Myr (solid), in an accretionally heated disk. 
Water content with heliocentric distance is seen to change significantly with the diffusion treatment used, as both 
GH06/CC06 and SL88 predict faster volatile diffusion timescales than our work, and therefore show greater enhancement of 
icy-rocky material (in comparison to initial $\Sigma_{\rm rocky chondrules}$ shown by black dashed line) just beyond the 
snowline, between 1-3 AU. 
We note that the greatest deviation from our work is with the GH06/CC06 profiles that show a sustained enhancement in ice-to-rock 
ratio by a factor of 2 just beyond the snow line at 0.1 and 1 Myr. 
As timescales for core growth are inversely proportional to solids-to-gas ratio (Kokubo \& Ida 2002), an erroneous enhancement 
of icy and rocky material would seem to decrease core growth timescales significantly, overestimating rate of growth of planetesimals 
and eventually planets.

%-----------------------------------------------------------------------------------------------------------
%
% FIGURE 4
%
\begin{figure}[ht!]
\begin{center}
\includegraphics[width=0.50\paperwidth]{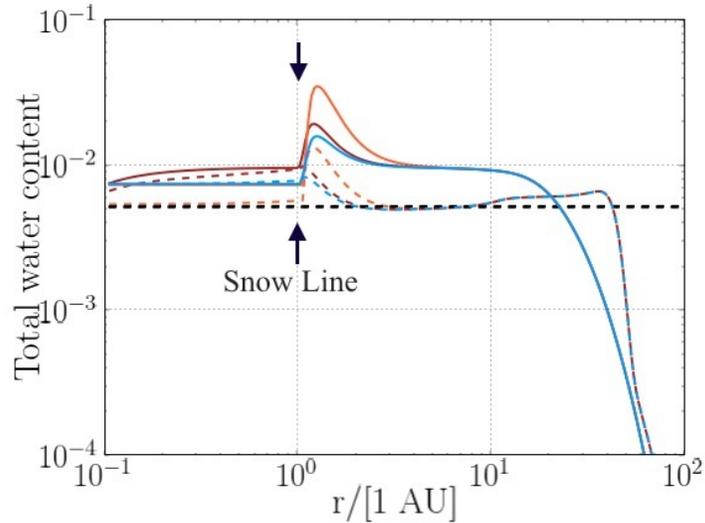}
% \plotone{Fig4.jpg}
\caption{Simulated evolution over time of the total water abundance across disk radius $r$, i.e., ($\Sigma_{\rm vapor} + \Sigma_{\rm icy chondrules} + \Sigma_{\rm icy asteroids} / \Sigma_{\rm gas}$) in a disk with accretional heating. Same as Figure 1: Blue profiles show 0.1 Myr (dashed) and 1 Myr (solid) with the formulation used here (cf. Gail 2001). Orange profiles similarly show formulations used by GH06/CC06, and brown profiles by SL88.} 
\end{center}
\end{figure}
%-----------------------------------------------------------------------------------------------------------

\section{Summary}

In the current literature, nine independent derivations have resulted in six mutually exclusive equations for volatile transport.
These different treatments make significantly different predictions about the abundance of water and the surface density of ice.
The large difference in predicted outcomes underscores how important it is to use the correct equation to
calculate radial transport of volatiles. 
We have derived the volatile transport equations starting with Fick's law and have identified the correct equations to use.
With the discrepancies between existing treatments explained and the correct forms identified, we hope that this
paper can serve as a resource for the disk modeling community. 

\acknowledgments 
\vspace{0.3in}
The authors tried to review the literature as comprehensively as possible, and apologize if they overlooked other 
original treatments of radial diffusion in protoplanetary disks.

This work is partially supported by a grant from the Keck Institute for Space Studies.

%\begin{center}
%{\bf References} 
%\end{center}

\end{document}